\documentclass[journal=ACSNANO,manuscript=article]{achemso}

\usepackage[version=3]{mhchem} 
\usepackage{color}
\usepackage{gensymb}
\usepackage{fancyhdr}

\newcommand{\comZ}[1]{{{#1}}}   
\author{Ricardo Javier Peña Román}
\email{rikrdopr@ifi.unicamp.br}
\author{Fábio J. R. Costa }
\affiliation[Unicamp]
{Institute of Physics ``Gleb Wataghin'', Department of Applied Physics, State University of Campinas-UNICAMP, 13083-859, Campinas, Brazil}
\author{Alberto Zobelli}
\affiliation[LPS]
{Universit\'e Paris-Saclay, CNRS, Laboratoire de Physiques des Solides, 91405, Orsay, France}
\author{Christine Elias}
\author{Pierre Valvin}
\author{Guillaume Cassabois}
\affiliation[LCC]
{Laboratoire Charles Coulomb, UMR5221 CNRS-Université de Montpellier, 34095 Montpellier, France}
\author{Bernard Gil}
\affiliation[LCC]
{Laboratoire Charles Coulomb, UMR5221 CNRS-Université de Montpellier, 34095 Montpellier, France}
\author{Alex Summerfield}
\author{Tin S. Cheng}
\author{Christopher J. Mellor}
\author{Peter H. Beton}
\author{Sergei V. Novikov}
\affiliation[SPA]{
School of Physics and Astronomy, University of Nottingham, Nottingham, NG7 2RD, UK}
\author{Luiz F. Zagonel}
\email{zagonel@unicamp.br}
\affiliation[Unicamp]
{Institute of Physics ``Gleb Wataghin'', Department of Applied Physics, State University of Campinas-UNICAMP, 13083-859, Campinas, Brazil}

\title[An \textsf{achemso} demo]
{Band gap measurements of monolayer h-BN and insights into carbon-related point defects}

\abbreviations{h-BN, STM, STS, PL, CL, HOPG}

\begin{document}



This is the version of the article before peer review or editing, as submitted by an author to IOP 2D Materials.

Find the published version at:
https://doi.org/10.1088/2053-1583/ac0d9c

Cite as: Ricardo Javier Peña Román et al 2021 2D Mater. 8 044001

\newpage

\begin{abstract}

Being a flexible wide band gap semiconductor, hexagonal boron nitride (h-BN) has great potential for technological applications like efficient deep ultraviolet (DUV) light sources, building block for two-dimensional heterostructures and room temperature single photon emitters in the UV and visible spectral range. To enable such applications, it is mandatory to reach a better understanding of the electronic and optical properties of h-BN  and the impact of various structural defects. Despite the large efforts in the last years, aspects such as the electronic band gap value, the exciton binding energy and the effect of point defects remained elusive, particularly when considering a single monolayer.

Here, we directly measured the density of states of a single monolayer of h-BN  epitaxially grown on highly oriented pyrolytic graphite, by performing low temperature scanning tunneling microscopy (LT-STM) and spectroscopy (STS). The observed h-BN electronic band gap on defect-free regions is $(6.8\pm0.2)$ eV. Using optical spectroscopy to obtain the h-BN optical band gap, the exciton binding energy is determined as being of $(0.7\pm0.2)$ eV. In addition, the locally excited cathodoluminescence and photoluminescence show complex spectra that are typically associated to intragap states related to carbon defects. Moreover, in some regions of the monolayer h-BN we identify, using STM, point defects which have intragap electronic levels around 2.0 eV below the Fermi level.

\end{abstract}

\section{Introduction}

Hexagonal boron nitride (h-BN) is a layered compound that is isomorphous with graphite.  In its bulk form, h-BN is formed from monolayers composed of boron and nitrogen atoms in a hexagonal \emph{sp}$^2$ covalent lattice that are organized vertically by van der Waals (vdW) interactions.\cite{Adachi1999,Alem2009,BHIMANAPATI2016} With an optical band gap of about 6 eV, and an indirect-to-direct band gap crossover in the transition from bulk to monolayer, \cite{watanabe2004,Cassabois2016,Elias2019} h-BN shows very bright deep ultraviolet (DUV) emission \cite{WATANABE2011,Kubota2007,Pierre2020} and defect mediated emission from the DUV all the way to the near-infrared.\cite{Museur2008,Bourrellier2014,Schu2016,Vuong2016,Gil2020}. In particular, point defects have been observed to act as single-photon sources.\cite{Tran2015,Bourrellier2016,Tran2016, Tran2016-2,Xu2018,Tan2019,Hayee2020,Mendelson2020,Mart2016} Such properties place bulk h-BN and its monolayer form in the spotlight for many potential applications, including DUV light emitting devices,\cite{watanabe2009,Watanabe-UV2011,Jiang2014} dielectric layers for two-dimensional (2D) heterostructures\cite{Hui2016,Wang2017,Wu2020} and room temperature (RT) single photon emitters (SPEs) for quantum technologies.\cite{Wrachtrup2016,Aharonovich2016,Aharonovich2017,Kim2019,Lukishova2020} Similarly to other technologically-relevant semiconductors, the successful application of h-BN depends on the understanding and control of its electronic and optical properties. However, partially due to its wide band gap and challenging large area synthesis, many fundamental electronic and optical properties of h-BN remain elusive. Moreover, the morphology, the electronic properties and the optical emission of structural point defects are complex and are currently poorly understood.

Given the wide band gap of h-BN and its insulating character, studying its morphological, electronic and optical properties can be tricky, particularly for the case of a single monolayer. The band structure of bulk and monolayer h-BN have been reported by Angle Resolved Photoemission Spectroscopy (ARPES).\cite{Hecnk2017, Pierucci2018} However, ARPES cannot resolve the conduction band structure. Electron Energy Loss spectroscopy (EELS) has already been applied to measure the band gap on h-BN nanotubes and h-BN monolayers, but EELS measures the optical band gap, similarly to optical absorption.\cite{Arenal2005,Liu2015} 
Therefore, scanning tunneling spectroscopy (STS) becomes an appropriate approach to probe both the valence band and conduction band edges as well as for the determination of the electronic band gap.\cite{FEENSTRA1994-,Feenstra1994} Still, so far, STS has not provided a clear measurement of the electronic band gap for monolayer h-BN or intra-gap states related to defects. The structure of individual defects in h-BN has been explored in previous works by means of scanning tunneling microscopy (STM)\cite{Wong2015} and high resolution transmission electron microscopy,\cite{Jin2009} but, despite these efforts, a clear correlation between electronic levels and light emission related to defects has not been established yet. Optically, the properties of few-layer h-BN samples have recently been investigated by cathodoluminescence (CL) measurements. However, the thinnest h-BN samples for which a CL signal could be collected were six \cite{Schue2016} and three\cite{Hernandez2018} monolayers flakes. Consequently, the electronic band gap value and the exciton binding energy are unaddressed experimentally even now, alongside the CL emission in the monolayer and electronic signatures of defect states. Finally, the direct character of the band gap and optical transitions in the DUV in monolayer h-BN have been recently reported\cite{Elias2019}, but the optical transition in the near UV and visible spectral range have not yet been extensively explored. 

Here, we show that monolayer h-BN epitaxially grown on highly oriented pyrolytic graphite (HOPG) is a model system for the study of the morphological, electronic,  and optical properties of h-BN monolayers. The weak sample-substrate interaction enables, for the first time, the determination of the electronic band gap value by means of STS measurements. Correlation between tunneling spectroscopy with DUV optical spectroscopy enables a value for the exciton binding energy in monolayer h-BN to be estimated. Additionally, structural point defects are observed by STM images. Luminescence due to defects in a broad emission range is observed for the first time using CL in monolayer h-BN. Moreover, insights on the optical signatures of defects are obtained from the interesting contrast between \emph{in situ} photoluminescence (PL) and CL. These results indicate that monolayer h-BN does not form significant interface states with HOPG, so that h-BN exhibits its fundamental electronic and optical properties.

\section{Results and discussion}

\subsection{\label{Sec:General} Sample and methods descriptions}

\begin{figure}
    \centering
\includegraphics[width=6.3 true in]{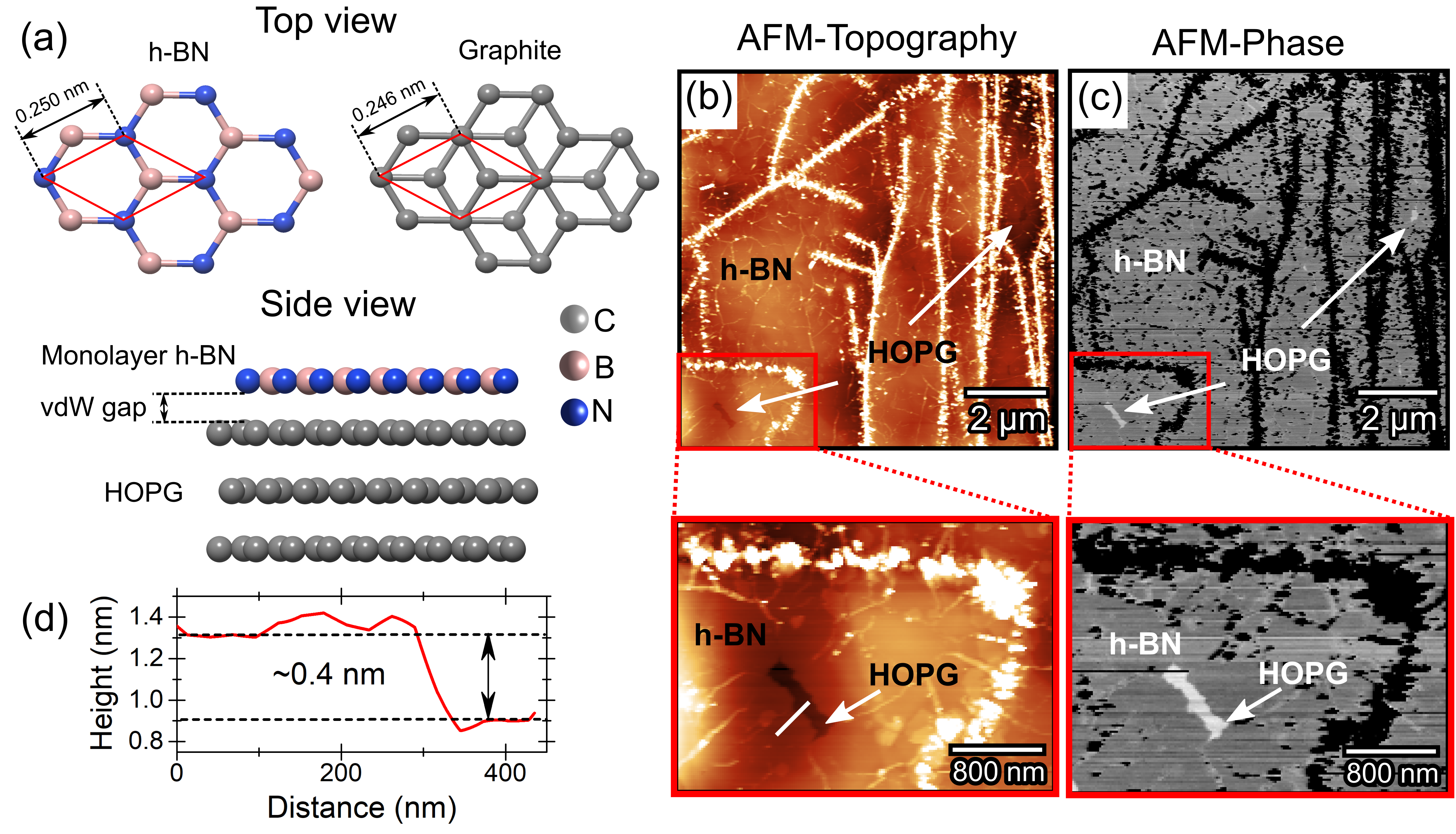}
\caption{\label{Fig:Sample} (a) Top view of the in-plane atomic structure of h-BN and HOPG. Side view of the monolayer h-BN/HOPG van der Waals heterostructure. AFM (b) topographic and (c) phase images of the epitaxial monolayer h-BN/HOPG. (d) Height profile of a single monolayer h-BN. }
\end{figure}

In Figure~\ref{Fig:Sample}, a general description of the h-BN sample investigated in this work is displayed. As shown in Figure~\ref{Fig:Sample}(a), h-BN and HOPG have an in-plane hexagonal structure with similar lattice parameters. The small lattice mismatch makes them highly compatible and appropriate for the epitaxial growth of vertical heterostructures. Additionally, the surfaces of these kinds of materials are naturally passivated without any dangling bonds. Therefore, in vdW epitaxial heterostructures, an atomically sharp interface is obtained, where chemical bonds are absent and  only vdW interactions are present between the sample and the substrate.\cite{Saenz1994,KOMA1999,Novoselov2016,WALSH2017,Choi2017} The region at the interface is usually referred to as vdW gap, such as illustrated in Figure~\ref{Fig:Sample}(a). Since the vdW interaction is weak, it is expected that each material in the heterostructure preserves most of its electronic properties. This was revealed in recent works by ARPES measurements. Sediri \emph{et al.}\cite{sediri2015} proved that when a monolayer h-BN is epitaxially grown on graphene (Gr), the electronic structure of Gr remains unaffected under the presence of h-BN. In the same way, Pierucci \emph{et al.} \cite{Pierucci2018} demonstrated in monolayer h-BN/HOPG that the electronic properties of h-BN are not perturbed significantly by the substrate. This means that in vdW heterostructures, the sample and the substrate are electronically decoupled, and there are no doping effects or charge transfer. All these effects make the vdW heterostructure of monolayer h-BN on HOPG an important model system for the study of the fundamental properties of h-BN, and for applications when conductivity is relevant as in STM/STS and light emitting diodes.   

 The h-BN layers were grown on HOPG substrates by high-temperature plasma-assisted molecular beam epitaxy (PA-MBE).\cite{Elias2019, Summerfield2018, Vuong2017,Cho2016} Figures~\ref{Fig:Sample}(b) and (c) show large area topographic and phase atomic force microscopy (AFM) images, respectively, of the h-BN layer on the HOPG substrate, acquired with the amplitude-modulated tapping mode in ambient conditions. As observed, the HOPG surface is almost totally covered by h-BN, where the white regions in the phase image correspond to uncovered HOPG surface areas. The AFM topographic image in Figure~\ref{Fig:Sample}(b) shows  that the sample is predominantly composed by large terraces of monolayer h-BN with some small bilayer regions and some thick regions around the HOPG grain boundaries and surface steps, which appear as bright (dark) features in the topography (phase) zoomed image (see also Figure S1). The height profile in Figure~\ref{Fig:Sample}(d), taken along the white line on the h-BN/HOPG surface of Figure~\ref{Fig:Sample}(b), shows that the step height is approximately 0.4 nm, in agreement with the expected thickness of a single monolayer h-BN. This morphology is consistent with previous reported results on similar samples. \cite{Summerfield2018,Vuong2017,Elias2019,Cho2016}.

\begin{figure}[h!]
    \centering
\includegraphics[width=6.3 true in]{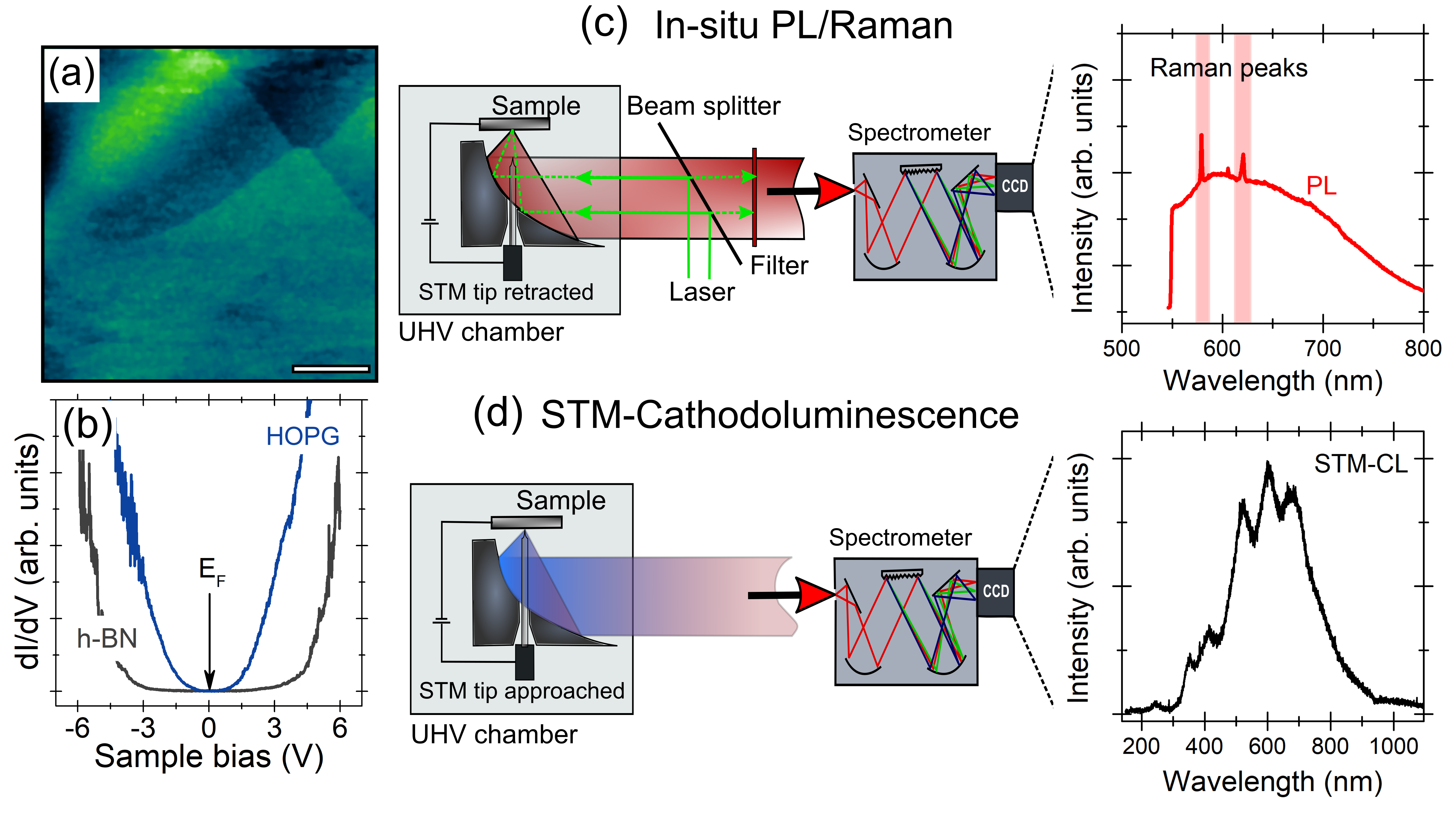}
\caption{\label{Fig:General_scheme} (a) STM image of the monolayer h-BN/HOPG surface (0.8 nA, 0.2 V, 80 K). Scale bar of 30 nm. (b) STS curves at 80 K (The feedback loop was disabled at
-6.2 V, -150 pA.). Schematic illustration of the experimental setup for (c) \emph{in-situ} PL/Raman and (d) STM-CL measurements in the STM configuration. The spectra shown here are uncorrected raw data.}
\end{figure}

The surface morphology and electronic properties of the  monolayer h-BN were investigated locally by means of STM and STS measurements performed under ultra-high vacuum (UHV) conditions at cryogenic temperatures using a modified RHK PanScan FlowCryo microscope. An STM image revealing the smooth sample surface is shown in Figure~\ref{Fig:General_scheme}(a), as observed previously by AFM (Figure S1). Considering only the STM image, it is difficult to know if the imaged area corresponds to the h-BN or to the HOPG surface because they are morphologically very similar. Despite being isomorphic, h-BN and HOPG have striking different electronic structures (wide band gap semiconductor and zero-gap conductor, respectively) and therefore very different local density of states (LDOS) to be measured by STS. The gray curve in Figure~\ref{Fig:General_scheme}(b) is a typical STS spectrum obtained in the region of Figure~\ref{Fig:General_scheme}(a). This dI/dV curve shows an electronic band gap of $\sim$6 eV, defined by the low differential conductance range from $\sim$-3 V to $\sim$+3V, confirming that the scanned region corresponds to a h-BN covered surface. \comZ{Considering the general morphology of the h-BN grown on HOPG observed by AFM and the fact the the majority of the sample is covered by monolayers, the h-BN thickness in smooth regions as that one in Figure ~\ref{Fig:General_scheme}(a) is a considered to be a monolayer.} In other regions with similar surface morphology, the STS curves exhibit the characteristic parabolic shape around the Fermi level of the LDOS of graphite,\cite{zhou2006,Castellanos2012} as in the blue plot in Figure~\ref{Fig:General_scheme}(b). 

 The STM was adapted to include a high numerical aperture light collection and injection system with optimized transmission. This system uses an off-axis parabolic mirror inserted around the tunneling junction and  \emph{in situ} PL and Raman spectroscopies can be performed using the light injection and collection system as illustrated in Figure~\ref{Fig:General_scheme}(c). In this experimental configuration, the STM tip is retracted from the sample surface and the light of a green (532 nm) laser diode is injected into the STM junction under UHV conditions. Additionally, STM-induced light emission or the CL response of the sample can be investigated using the setup, see Figure~\ref{Fig:General_scheme}(d). For recording the CL signal, the STM is operated in field emission mode using an external voltage source of (0-500) V for the local excitation of the sample by electron bombardment (STM-CL). The PL/Raman and CL spectra, shown in Figure~\ref{Fig:General_scheme}(c) and (d), correspond to the spectroscopic raw data that need to be corrected for the proper interpretation, as explained in the Supplementary materials (SM). This setup has also been previously used to investigate WSe$_2$ and MoS$_2$ monolayers.\cite{D0NR03400B,DOAMARAL2021}.

\subsection{\label{Sec:STM/STS_LT} Electronic band gap, optical band gap and exciton binding energy}

Since monolayer h-BN is an atomically thin and a wide band gap material, access to its electronic structure is always a highly challenging experimental and theoretical task. Within the last twenty years, the h-BN electronic band gap has been calculated a number of times within the GW approximation, while the optical response has been calculated using the Bethe-Salpeter equation \cite{Blase1995, Arnaud2006, Wirtz2006, Paleari2018}. In this approximation, the Green's function for the quasi-particles are evaluated within the screened Coulomb potential in a way to consider many-body problems of the interacting electrons.\cite{Aryasetiawan1998} However, non self-consistent GW calculations (G$_0$W$_0$), which start from DFT orbitals and perform the calculation only once, typically provide underevaluated values for the electronic band gap and, in the case of h-BN, a rigid shift of the optical spectra is required for a direct comparison with experiments. Only very recently, self-consistent GW calculations, which update interactively both the wave functions and the eigenvalues, managed a good agreement with synchrotron ellipsometry experiments for bulk h-BN.\cite{Arts2021} In the case of free-standing monolayer h-BN and using the GW$_0$ approach, which updates interactively the eigenvalues only in the Green's function, the electronic band gap has been evaluated as 8.2 eV and the related optical gap at 6.1 eV.\cite{Wirtz2006,Hunt2020} Slightly higher values might be expected within the full self-consistent GW approach. On the other hand, at the moment, there is no experimental evidence with effective measurements of the electronic band gap in both bulk or monolayer h-BN. A particular reason for the lack of such results is the technical challenge regarding the accomplishment of experiments for DUV wavelengths. The electronic properties and band structure of epitaxial monolayer h-BN on graphite have been studied recently by ARPES measurements.\cite{Pierucci2018}  This study reported important information about band alignment, Fermi level and Valence Band (VB) positions. However, it is noteworthy that ARPES measurements only resolve filled states,\cite{Yang2018,zhang2014,Hecnk2017,Pierucci2018} and therefore the description of the whole band structure and electronic band gap of monolayer h-BN are still open questions.

One technique able to probe both  filled and empty states, with a complete information about the LDOS, doping effects and charge transfer is scanning tunneling spectroscopy or STS.\cite{Feenstra1994,  Zang-2014,ugeda_giant_2014,huang2015,Zhang-2015,KRANE2018} Previous STM/STS measurements on h-BN have not given a clear answer for the electronic band gap. The main reason for that is because most of the STM/STS studies have been carried out on h-BN samples grown directly on metallic substrates.\cite{AUWARTER2019} Nowadays, it is well documented that the growth of a 2D material on a metallic substrate leads to a band gap renormalization due to strong interactions with the substrate, which is related to the dielectric screening by the metal and/or the formation of additional interface electronic states, including hybridization among others.\cite{Bruix2016,Dendzik2017,zhang_bandgap_2016,Pan2016,DOAMARAL2021} These effects reduce the electronic band gap value to about 3 eV for h-BN on Ru(0001)\cite{zhang_bandgap_2016} and on Re(0001)\cite{Qi2017} surfaces, to $\sim$4 eV for h-BN on Au(001),\cite{Zhang-2017} and to $\sim$5 eV for the case of h-BN on Rh(111),\cite{BRIHUEGA2008,Nataly2021} Ir(111),\cite{Liu2014} and Cu(111)\cite{Li2015} substrates. In addition, performing scanning tunneling experiments on h-BN can be complex due to its expected insulating character. For instance,  Wong, \emph{et al} used Gr as conducting layer on top of bulk h-BN.\cite{Wong2015} Using the Gr as conducting electrode affects considerably STM images and make STS spectra difficult to interpret, particularly with respect to the h-BN band gap. Interestingly, when h-BN is grown on Gr/Cu(111), electronic states of the metallic substrate are observed on the Gr and also on the h-BN layer.\cite{Pan2016} In this case, intragap states due to the copper support are measured by STS and the h-BN band gap is not observed. For vdW epitaxial heterostructures, the sample-substrate interaction is not an issue because, as mentioned above, there is a sharp interface between the monolayer h-BN and the HOPG substrate governed by weak vdW bonds (vdW gap). Therefore, h-BN on HOPG can be considered as being electronically decoupled from the substrate, and is not expected to demonstrate intragap states coming from graphite, but only effect of the dielectric environment surrounding the monolayer.

\begin{figure}[h!]
    \centering
\includegraphics[width=6.3true in]{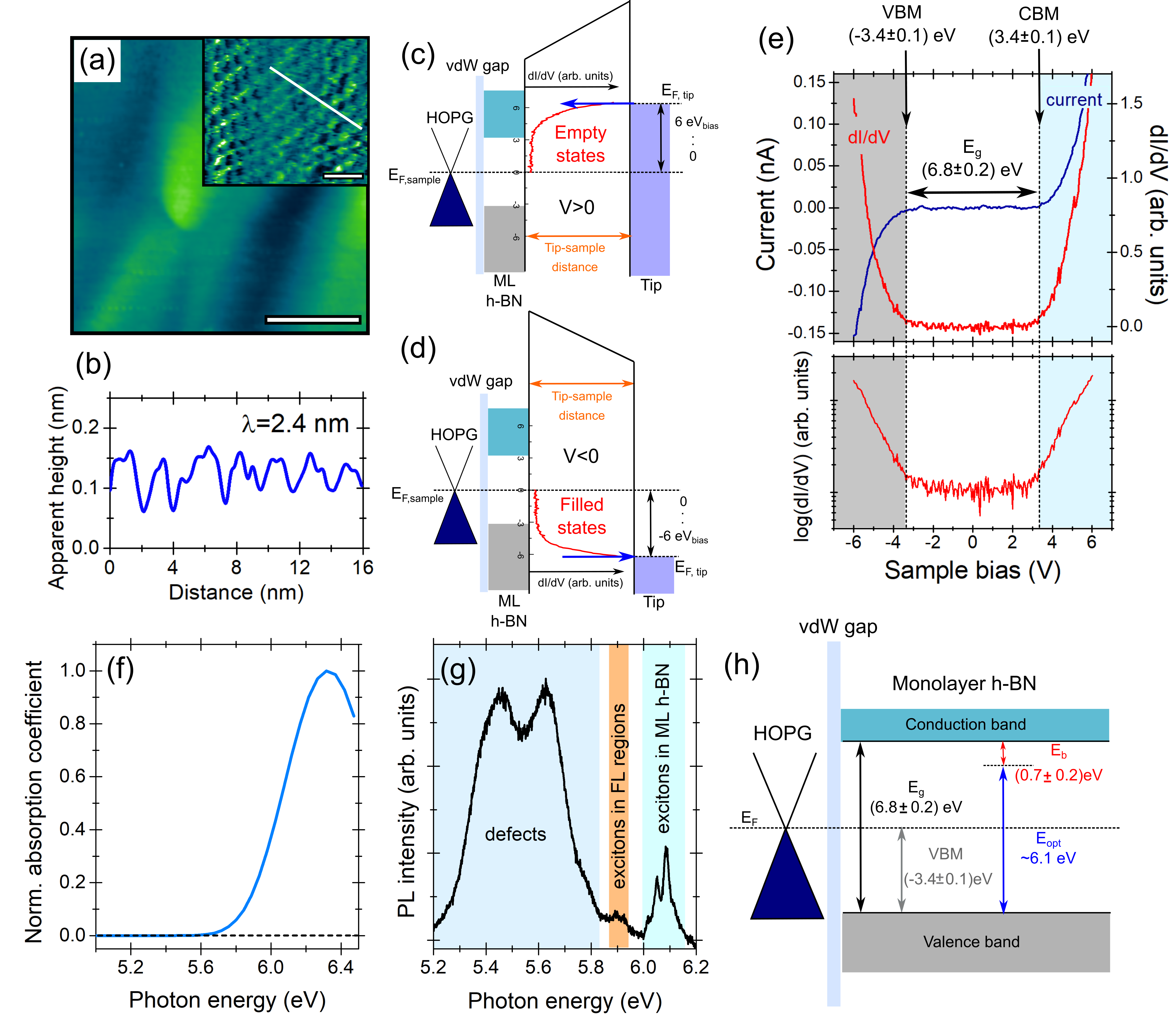}
\caption{\label{Fig:STM_STS} (a) STM image of a smooth and defect-free monolayer h-BN region (0.66 nA, 3.8 V, 80 K). Scale bar of 30 nm. Inset: moiré structure (-0.3 nA, -1.0 V, 16 K). Scale bar of 5 nm. (b) Height profile associated to the moiré. Schematic illustration of the tunneling process for (c) positive and (d) negative biases. The red curve corresponds to the LDOS measured by STS as dI/dV. The blue arrow indicates the tunneling direction. (e) I-V and dI/dV typical curves obtained on a smooth and defect-free monolayer h-BN region and at 80 K (The feedback loop was disabled at -6.2 V, -150 pA). The electronic band gap and band edges are indicated with the values obtained of a statistical analysis of 480 individual curves. Global measurements of the (f) optical absorption at RT and (g) PL at  10 K. (h) Energy levels diagram proposed for defect-free monolayer h-BN. In the figure VBM, E$_\text{g}$, E$_\text{opt}$ and E$_\text{b}$ correspond to the valence band maximum, the electronic band gap, the optical band gap and the exciton binding energy, respectively.}
\end{figure}


 Figure~\ref{Fig:STM_STS}(a) and (b) shows STM and STS results acquired at 80 K in the monolayer h-BN/HOPG. A STM image of a smooth and defect-free region of the sample surface together with an insert showing a moiré structure with with a periodicity of approximately 2.4 nm are displayed in Figure~\ref{Fig:STM_STS}(a). This moiré pattern corresponds to a local rotation of approximately 5.8° between the monolayer h-BN and the HOPG surface in a particular region, as recently observed on similar samples.\cite{Summerfield2018} The apparent height profile of the moiré is presented in Figure~\ref{Fig:STM_STS}(b). Detailed studies of the moiré structure in this system were reported in recent works. \cite{Summerfield2018,Thomas2020} 
 
 The LDOS of the monolayer h-BN has been investigated by performing STS measurements on the defect-free region at different tip positions. The tunneling processes that probe both empty and filled states by STS in the monolayer h-BN are illustrated in Figure~\ref{Fig:STM_STS}(c) and (d), respectively. The alignment between the Fermi level in h-BN and HOPG at 0 V was demonstrated in the results depicted in Figure~\ref{Fig:General_scheme}(b). Thus, when a positive sample bias ranging from 0 V to +6 V is applied, the Fermi level in the tip is shifted with respect to the Fermi level in h-BN/HOPG, and, as a consequence, electrons tunnel from the tip to empty states of the sample, such as shown in Figure~\ref{Fig:STM_STS}(c). The red curve in the figure represents the measured dI/dV for positive biases, which is a direct measurement of the local empty DOS. It can be observed that the LDOS increases from biases above $\sim$3.0 V, which means that electrons are tunneling only for states in the Conduction Band (CB) of h-BN. Below $\sim$3.0 V the curve is flat (the tunneling current is equal to $\sim$0 pA), because there are no available electronic states inside the h-BN electronic band gap for electrons to tunnel to. The same process happens when the polarity of the bias is changed to negative values in order to probe filled states, as illustrated in Figure~\ref{Fig:STM_STS}(d). It is worthwhile to note that, during typical STS measurements, the voltage ramps while the current is measured but the tip-sample distance is not adjusted. Therefore, the tip-sample distance is set by the tunneling current and sample bias prior to disabling the tip-sample distance feedback loop and starting the STS measurement (stable tunneling parameters). Here, we used typically -150 pA at -6.2 V (see Methods section for more details) as the stable condition prior to starting the acquisition of the scanning tunneling spectra in order to ensure reproducible and reliable spectra. If the tunneling current is set to slightly higher or lower values (from -50 to -300 pA) before disabling the feedback loop, the obtained spectra are compatible to those measured using -150 pA, see Figure S2.
 
 It is relevant to mention that direct tunneling from or to the HOPG is possible even through the monolayer h-BN. In fact, this effect can be observed when STS curves were recorded stabilizing the tunneling junction at high tunneling currents ($\geq$ 600 pA) with -6.2 V of sample bias, as discussed in the SM, see Figure S2. A possible explanation for this observation is that for a fixed sample bias and higher stabilization currents, the tip-sample distance becomes shorter, which means that the tunneling barrier is narrower, and then the probability of the direct tunneling from or to the substrate is not negligible. This leads to the measurement of an apparent reduced band gap, because states of the HOPG near the band edges of h-BN are observed in the STS curve for h-BN, as can be seen in Figure S2(b). The direct tunneling from/to the HOPG substrate for short tip-sample distance in STS measurements on vdW epitaxy MoS$_2$/HOPG and WSe$_2$/HOPG samples have been demonstrated by Chiu \emph{et al.}\cite{chiu_determination_2015} In that work, it was also demonstrated that the LDOS of HOPG can be measured by tunneling electrons through the TMD layer if the stabilizing bias for the STS acquisition lays inside the band gap region of the TMD. For wide band gap materials such as h-BN, using appropriate tunnel junction conditions before starting the STS is therefore crucial.

Figure~\ref{Fig:STM_STS}(e) shows typical STS results obtained from a smooth and defect-free h-BN sample surface region. The top panel of the figure shows the I-V curve plotted together with the dI/dV tunneling spectrum. The electronic band gap of the h-BN sample is defined by the bias range at which the current and dI/dV are close to their background values. The band edges and band gap values were extracted from the dI/dV curve in logarithmic scale,\cite{ugeda_giant_2014,Bradley2015} as represented in the bottom panel of Figure~\ref{Fig:STM_STS}(e). The values shown in the figure were obtained following the method and the statistical analysis explained in the SM, Figures S3 and S4, which considers a total number of 480 individual dI/dV curves similar to Figure~\ref{Fig:STM_STS}(e) taken at different tip positions. The final results are a Valence Band Maximum (VBM) and a Conduction Band Minimum (CBM) positioned at $(-3.4\pm0.1)$ eV and  $(3.4\pm0.1)$ eV, respectively, for monolayer h-BN on HOPG. This results in an electronic band gap of $(6.8\pm0.2)$ eV. Also, the Fermi level is in the middle of the band gap, indicating that the defect-free monolayer h-BN regions, as shown in Figure ~\ref{Fig:STM_STS}(a), are undoped and therefore there is no charge transfer with the HOPG. ARPES found the VBM position close to the $\sim$ -2.8 eV on a similar h-BN sample.\cite{Pierucci2018} This ARPES result for the VBM is slightly different than the value measured by STS (-3.4 eV). This difference is possibly caused by some degree of p-type doping on the sample and/or some level of hole excess coming from a insufficient compensation of the photo-emitted electrons. On 3-monolayer thick h-BN samples, ARPES determined the VBM at -3.2 eV, which, considering the band structure for 3-monolayer thick h-BN and the expected electronic band gap opening for a monolayer h-BN, agrees well with the results presented here. \cite{Zribi2020} From the experimental point of view, STS is not sensitive to  the direct or indirect nature of the electronic band gap. Nonetheless, recently, it was confirmed the recombination of direct excitons in monolayer h-BN,\cite{Elias2019} which suggests that this is a direct band gap material.

The electronic band gap value found here for monolayer h-BN is larger, as expected from simulations,\cite{Paleari2018} than values for bulk single crystal and thin films of h-BN, as obtained indirectly by photocurrent spectroscopy, which results in an electronic band gap of about 6.45 eV.\cite{Musser2011,Uddin2017} Regarding theoretical predictions for free-standing monolayer h-BN, GW$_0$ calculations indicate a band gap of 8.2 eV at the K point of the Brillouin zone with an exciton binding energy of 2.1 eV, giving an optical band gap of 6.1 eV.\cite{Wirtz2006, Hunt2020} This value is close to the optical onset of 6.3 eV measured by EELS on a free-standing monolayer.\cite{Liu2015}  Simulations for monolayer h-BN  on top of graphene have shown a strong renormalization of the electronic band gap, of the order of 1 eV, due to strong screening effects of the substrate.\cite{Huser2013,Guo2021} The simulations also show that the electronic band gap renormalization of monolayer h-BN is basically the same on top of graphene and on top of graphite.\cite{Huser2013} This last scenario is the perfect model for our system of monolayer h-BN on HOPG and gives a strong support for the band gap value measured here by STS.


Optical absorption and PL were performed in the DUV to determine the optical (excitonic) band gap of monolayer h-BN on HOPG (see details in SM). Figure~\ref{Fig:STM_STS}(f) and (g) show measurements of the optical absorption and PL (spot-size of 200 $\mu$m and 50 $\mu$m, respectively) obtained at RT and at 10 K, respectively. The optical absorption spectrum was calculated from spectroscopic ellipsometry data. \comZ{The PL in the DUV employs a 6.4 eV laser obtained from the fourth harmonic of a Ti:Sa oscillator.} A systematic study on the complex DUV absorption and emission of high-temperature PA-MBE h-BN on HOPG can be found in recent reports.\cite{Elias2019,Vuong2017} The spectrum in Figure~\ref{Fig:STM_STS}(f) shows an abrupt change in the absorption coefficient that suggests an optical band gap of above 5.7 eV. Furthermore, the PL in  Figure~\ref{Fig:STM_STS}(g) shows the direct fundamental exciton transition at 6.1 eV (optical band gap) in monolayer h-BN, besides other emission lines associated with transitions in few-layer and defective regions around 5.9 eV and 5.6 eV, respectively. These results are in excellent agreement with previous experiments by Elias \emph{et al.}\cite{Elias2019}  Moreover, this value for the optical band gap is consistent with recent simulation and indicate, as expected, an invariance of the exciton emission.\cite{Wirtz2006, Hunt2020, Guo2021, ugeda_giant_2014}
 


The knowledge of both the electronic band gap and the optical band gap is essential to the design of applications in optoelectronics and light emitting devices employing large exciton binding energy materials, like the monolayer h-BN. Considering the h-BN electronic band gap found by STS and the energy observed for the exciton recombination, the obtained exciton binding energy is $(0.7\pm0.2)$ eV. The comparison between this experimental evaluation and the theoretical predicted value of 2.1 eV for the free-standing monolayer, suggest a strong renormalization, of about 1.4 eV, of the exciton binding energy due to substrate screening effects.\cite{Hunt2020} Moreover, the observed renormalization of the exciton binding energy is expected to be the same on the electronic band gap, which nearly leads to an invariance of the optical onset.\cite{Guo2021,ugeda_giant_2014}
This renormalization is in close agreement with simulations for monolayer h-BN on top of graphite.\cite{Huser2013} The similar shift on the electronic band gap and on the exciton biding energy agrees well with simulations for monolayer h-BN and with experiments and simulations for monolayer MoSe$_2$.\cite{Guo2021,ugeda_giant_2014} Based on all the spectroscopic data, Figure~\ref{Fig:STM_STS}(h) summarizes the energy levels determined here for a defect-free region of monolayer h-BN on HOPG. In Figure~\ref{Fig:STM_STS}(h), the h-BN electronic band gap is indicated as $(6.8\pm0.2)$ eV, while the h-BN optical band gap is 6.1 eV. This leads to a Frenkel exciton with a binding energy estimated as $(0.7\pm0.2)$ eV.


\subsection{\label{Sec:Defects} Electronic Structure and Light Emission related to  point defects}

The wide electronic band gap in h-BN allows the observation of several optical transitions involving different intragap states associated with defects.\cite{Museur2008,Schu2016,Vuong2016,Sajid_2020} Therefore, light emission in h-BN may be dominated by structural defects.  Point defects have a particular interest due to their characteristic quantum emission at RT in a broad range of wavelengths.\cite{Tran2016,Bourrellier2016,Tran2016-2,Tan2019} Different types of point defects have been proposed as being the sources of the single photon emission observed in h-BN.\cite{Gil2020,Sajid_2020,Zhang2020} However, their morphological and electronic signatures have not been measured directly. Furthermore, a direct correlation with the optical emission is missing. Even though STM is a tool able to identify atomic defects in 2D materials, \cite{D0NR03400B,Edelberg2019,Addou2015,Ziatdinoveaaw2019,zhussupbekov2021}
STM/STS measurements of point defects in monolayer h-BN have not been reported yet. The properties of individual point defects have been investigated only by simulated STM images in recent theoretical works.\cite{Fujimoto2016,fujimoto_formation_2017,Haga_2019,Haga-2019} Here some results obtained on defective regions of the sample are presented in order to help acquiring a better understanding of the impact of point defects in the properties of monolayer h-BN.

\begin{figure}[h!]
    \centering
\includegraphics[width=6.3 true in]{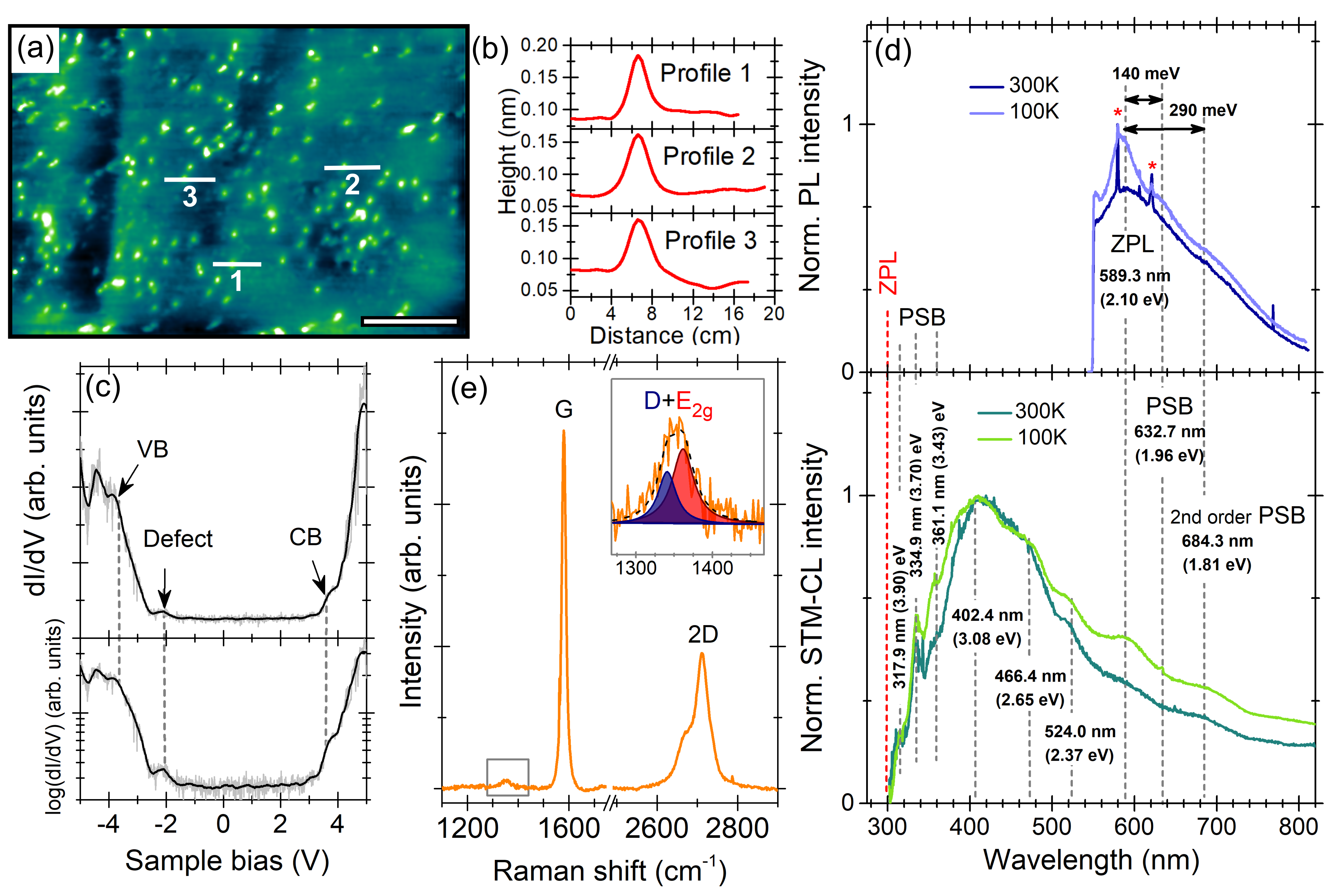}
\caption{\label{Fig:defects} (a) STM image of bright point defects in monolayer h-BN (0.8 nA, 0.8 V, 80 K). Scale bar of 30 nm. (b) Heigth profile of some individual defects. (c) Typical STS curve obtained on the defective region at 80 K (The feedback loop was disabled at -6.2 V, -150 pA). (d) \emph{In-situ} PL and STM-CL spectra in which Phonon Side-Bands (PSB) are observed near Zero Phonon Loss (ZPL) peaks. (e) \emph{In-situ} Raman spectrum at RT. Inset: Lorenztian fitting of the peak at 1350 cm$^{-1}$.}
\end{figure}

Figure~\ref{Fig:defects}(a) shows a STM image of the monolayer h-BN/HOPG surface, where point defects are revealed. These defects are bright spots of about 1 to 2 nm of diameter, according to the full width at half maximum in the height profiles of Figure~\ref{Fig:defects}(b), see also Figure S5. Defects in h-BN on HOPG were already observed by Summerfield \emph{et al.}\cite{Summerfield2018} using conductive AFM and STM measurements, and were attributed to defects in the HOPG substrate, created by the active nitrogen plasma irradiation damage during the sample growth process. In another study, bright point defects relating to possible carbon impurities were imaged by STM in Gr capped bulk h-BN.\cite{Wong2015} In order to inspect the electronic signature of the point defects found here, STS curves were acquired. Figure~\ref{Fig:defects}(c) displays a typical tunneling spectrum recorded when the STM tip is paced close to such a defect. By plotting the  STS curve in logarithmic scale, it can be noted that the bands onset are not linear, in contrast to the ones observed in defect-free regions (see Figure~\ref{Fig:STM_STS}(e) and Figure S3(c)). The curve in  Figure~\ref{Fig:defects}(c) presents three main resonances  at $\sim-3.5$ V, $\sim-2.0$ V and $\sim+3.5$ V. By comparing with the results for defect-free regions, the resonances at $\pm3.5$ V are located on top of the CB and VB edges, respectively, indicating that such defects have energy levels close to the band edges. Besides the spectra on Figure ~\ref{Fig:defects}(c), the electronic level near -2.0 V was observed in several other tunneling spectra obtained on regions with point defects, see Figure S5. Therefore, this energy level is attributed to the observed defects. The fact that the STS curves show roughly the expected band gap for h-BN even on defective regions suggests that the scanned regions corresponds to a h-BN covered area and the observed point defects are in the monolayer h-BN.

A good starting point to interpret the imaged defects and the STS results is to consider that in the system of monolayer h-BN on HOPG, the atoms of Boron (B), Nitrogen (N) and Carbon (C) are present. Thus, substitutional defects such as $\text{C}_\text{B}$ and  $\text{C}_\text{N}$, carbon dimer defects like $\text{C}_\text{B}\text{C}_\text{N}$, or carbon anti-sites vacancies   of  the types $\text{V}_\text{B}\text{C}_\text{N}$, and $\text{V}_\text{N}\text{C}_\text{B}$ are possible, and also are the main point defect candidates to be considered. Indeed, the random distribution of C atoms in the h-BN lattice is energetically possible,\cite{Jamroz2019} and on the other hand, the high temperature and plasma assisted growth process can create vacancies during the sample preparation.\cite{Summerfield2018,Xu2018} From the electronic point of view, band structure calculations show that for the LDOS in monolayer h-BN the states near the VBM are concentrated in the N sites, while the states near the CBM are in the B sites.\cite{Galvani2016,Jamroz2019} Impurities occupying N or B sites could act as acceptor-like or donor-like impurities, respectively. There are several theoretical reports exploring the doping effects  of C impurities in single layers of h-BN. In those works, it is demonstrated that  $\text{C}_\text{B}$ and  $\text{C}_\text{N}$ defects induce n-doping and p-doping, respectively.\cite{Fujimoto2016,Haga_2019,Berseneva2013, Attaccalite2011} In particular, $\text{C}_\text{N}$ defects introduce an acceptor state localized at 2.0 eV below the Fermi level and between 1-2 eV above the VBM.\cite{Fujimoto2016,Berseneva2013,Katzir1975} The density of this defect state depends on the C concentration, \cite{Jamroz2019} and the energy position  can be shifted due to local strain effects,\cite{Fujimoto2016,Sajid-2020,grosso_tunable_2017} and also by band bending effects.\cite{Edelberg2019} More recently, it also has been  predicted that the $\text{C}_\text{B}\text{C}_\text{N}$ dimer introduces a stable and neutral intragap state localized at $\sim$0.8 eV above the VBM.\cite{Mackoit2019,Jara2021}

It is important to point out that our STM results are consistent with the predictions made by Fujimoto \emph{et al.}\cite{Fujimoto2016} and by Haga \emph{et al.}\cite{Haga-2019,Haga_2019} in their works on simulated STM images of individual point defects in h-BN and Gr vertical heterostructures. In those reports, the  bright appearance of $\text{C}_\text{N}$ point defects were simulated for negative tunneling biases. The calculated STM images of C-doped monolayer h-BN show that C atoms produce a redistribution of the local electron density around the defects, being the electron density in $\text{C}_\text{N}$ higher than in $\text{C}_\text{B}$, which results in STM images where $\text{C}_\text{N}$ defects look as bright spots when compared with the image of $\text{C}_\text{B}$.\cite{Fujimoto2016,Haga_2019} Moreover, the local electron density, and as a consequence, the size of the bright spots related to $\text{C}_\text{N}$ defects can  be affected by the stacking and moiré pattern between the h-BN and Gr.\cite{Haga-2019,Haga_2019} Defects imaged with negative bias can be observed in Figure S5. These theoretical considerations indicate that the defects identified by STM are consistent with defects involving C on N sites, i. e.  $\text{C}_\text{N}$,  $\text{C}_\text{B}\text{C}_\text{N}$ and $\text{V}_\text{B}\text{C}_\text{N}$, even if other kinds of defect can not be excluded. Nonetheless, it is noteworthy that a detailed study of the morphological and electric structure of individual point defects in h-BN is required, which can be achieved by performing STM/AFM imaging and STS measurements at cryogenic temperatures, such as what has been done in the case of point defects in single layers of transition metal dichalcogenides.\cite{Zheng-2019,Schuler2019,barja_identifying_2019,Schuler2020}  

In order to obtain more insights on the observed defects, \emph{in situ} PL and STM-CL experiments were carried out at 100 K and 300 K, using the experimental setup described in  Figure~\ref{Fig:General_scheme}. The raw luminescence data were presented in Figure~\ref{Fig:General_scheme}(c) and (d), but for the proper interpretation, the spectra were corrected following the procedure explained in SM and the results are presented in Figure~\ref{Fig:defects}(d). The PL spectra show a main peak at 2.10 eV, which is best resolved at 100 K, and two shoulders around 1.96 eV and 1.81 eV. These transitions are related to the zero-phonon line (ZPL) and two phonon replicas or phonon side bands (PSB) of carbon-related defects, which behave as single photon sources.\cite{Mendelson2020} Simulations suggest  $\text{V}_\text{B}\text{C}_\text{N}$ or $\text{C}_\text{B}\text{V}_\text{N}$ as defects at the origin of this emission.\cite{Mendelson2020,Sajid-2020}  The width of the PL peaks is larger than that typically observed in thick h-BN flakes, but in agreement with reports on monolayers. \cite{Tran2015} Also in the PL spectra, two sharp peaks are observed and labeled with a red star symbol. These peaks correspond to the Raman response depicted fully in Figure~\ref{Fig:defects}(e), where the stronger peaks around 1580 cm$^{-1}$ and 2700 cm$^{-1}$ are the G and 2D Raman bands of HOPG, respectively.\cite{Zolyomi2011} In addition, an asymmetric and weak peak is observed at 1350 cm$^{-1}$, and as shown in the inset figure, this peak has two contributions. One contribution at 1341 cm$^{-1}$ associated to the D Raman mode of HOPG and another one at 1362 cm$^{-1}$ corresponding to the E$_{\text{2g}}$ Raman mode of monolayer h-BN, in agreement with recent reported results.\cite{Cho2016}

The CL spectra in the bottom panel of Figure~\ref{Fig:defects}(d) show a broad luminescence that ranges from the near infrared to the UV. As in PL, the peaks are better resolved at 100 K than at 300 K. Again, the optical transitions associated to carbon-related defects are observed with the  ZPL at 2.10 eV and the PSB at 1.81 eV, respectively, as seen in PL. At higher energies, the CL spectra present a series of peaks at 3.90 eV, 3.70 eV and 3.43 eV, which have been reported as being phonon replicas of a deep well-known carbon-related defect level at 4.1 eV.\cite{Pelini2019,Katzir1975} Unfortunately, the emission at 4.1 eV has not been resolved here due to the transmission function of the setup. The spectra only could be corrected for the instrument response function up to 300 nm (4.1 eV). However, the PSB were observed, which means that the transitions at 4.1 eV were also present in the emitted spectra. Interestingly, the light emission at 4.1 eV has been reported as being an optical transition related to the single photon emission.\cite{Bourrellier2016} Recent literature suggests that this emission is associated with $\text{C}_\text{B}\text{C}_\text{N}$ defects, carbon dimers, but other defects like $\text{C}_\text{N}$ are also considered in some cases. \cite{Mackoit2019, Jara2021, Weston2018,Linder2021} The most intense features observed in the CL spectra of Figure~\ref{Fig:defects}(d) are the peaks at 3.08, 2.65 and 2.37 eV, with shifts of 430 meV and 280 meV between them. Such peaks possibly have distinct origins related to the several possible defects, particularly in the presence of carbon.\cite{Weston2018} Emissions in this energy range have been observed in bulk h-BN by performing CL and PL experiments, and are associated to carbon\cite{Hayee2020, Katzir1975} impurities or nitrogen vacancy type-centers  \cite{Shevitski2019,BERZINA2016}.

\section{Conclusions}
We used low temperature UHV STM together with an optimized light collection and injection device to study the morphological, electronic, and optical properties of a monolayer h-BN epitaxially grown on HOPG. The STM images reveal h-BN regions free of defects and regions with point defects. An electronic band gap of $(6.8\pm0.2)$ eV was determined by performing STS measurements on defect-free monolayer h-BN. This result, combined with the h-BN optical band gap of 6.1 eV from the exciton transition, leads to a binding energy for the Frenkel exciton of $(0.7\pm0.2)$ eV. Additionally, bright point defects were observed in h-BN by STM imaging. The STS indicates an acceptor level around -2 eV related to the presence of the observed defects. PL and CL have shown the emission typically associated to carbon-related defects at 2.1 eV. Besides that, emissions at 3.08 eV and photon side band possibly associated to an emission at 4.1 eV were observed by CL on monolayer h-BN. These results indicate the simultaneous presence of more than one kind of carbon-related defect. We consider that the findings presented in this work could help in the understanding of the fundamental properties of monolayer h-BN, as well as in the identification at the atomic level of sources responsible for the SPEs in h-BN samples. Moreover, h-BN on HOPG represents an excellent platform to study individual defects with respect to their morphology, electronic and optical properties.

\section{Methods}

\subsubsection{Sample preparations}
Monolayer h-BN was grown on HOPG substrate by the high-temperature plasma-assisted molecular beam epitaxy (PA-MBE) method.\cite{Elias2019, Summerfield2018, Vuong2017,Cho2016} This sample preparation method allows to produce monolayer and few-layer h-BN with atomically flat surfaces and monolayer control of the sample thickness. The h-BN thickness and coverage can be controlled by substrate temperature, boron:nitrogen flux ratio and growth time. In particular, the sample investigated here was grown at a substrate temperature of about 1390 \degree C with a high-temperature effusion Knudsen cell for boron and a standard Veeco radio-frequency plasma source for active nitrogen. More details about the sample growth conditions and the MBE system can be found elsewhere.\cite{Cheng2018, Cheng2018-2,Cho2016,Wrigley2021}.

\subsubsection{STM/STS}
STM and STS measurements were performed under ultra-high vacuum (UHV) conditions at low-temperatures using a modified RHK PanScan FlowCryo microscope. This STM was adapted to receive a high numerical aperture light collection and injection system with optimized transmission. In this setup, imaging can be associated to electronic and optical spectroscopies as illustrated in Figure \ref{Fig:General_scheme}. 

Prior to STM/STS measurements, the h-BN sample was annealed at 773 K under a base pressure of 1.9x10$^{-10}$ mbar for about six hours, in order to eliminate surface impurities and contaminants. STM images were acquired in the constant-current mode using a grounded tungsten tip and a bias voltage applied to the h-BN sample. Tungsten tips were prepared by electrochemical etching with a NaOH solution. After preparation, the tips were inserted in the UHV chamber to be subjected to thermal annealing and argon sputtering. This procedure allows the removal tungsten oxide layers. For the acquisition of STS curves, differential conductance (dI/dV) measurements at 80 K were carried out using the lock-in technique, with a bias modulation of 80 mV of amplitude and 800 Hz of frequency.

To perform STS measurements, the tunnel junction was stabilized at a sample bias outside the electronic band gap of the monolayer h-BN. In most measurements, we stabilized the tip  at -150 pA and -6.2 V on a specific tip position before disabling the tip-sample distance feedback loop. Using these parameters, the tunnel current is effectively related to states on the h-BN. Each tunneling spectrum was recorded by turning off the tip distance feedback loop and ramping the bias from -6.0 V to 6.0 V in steps of 50 mV and a dwell time of 50 ms. This dwell time is needed to avoid overestimating the electronic band gap due to low signal to noise ratio near the band edges. A detailed description of the STS analysis is presented in the SM, Figures S3 and S4.

\subsubsection{\emph{In situ} PL, Raman and CL}
 
PL and Raman were excited with a laser diode of 532 nm. The spot size of the light focused by the mirror on the sample surface is $\sim2$ $\mu$m. The PL/Raman signal emitted by the sample is collected by the mirror and sent as parallel rays towards an optical setup outside the UHV STM chamber.
  
CL, also called STM-CL, was performed by operation of the STM in field emission mode. In this case, the STM tip was retracted by approximately 150 nm from the h-BN sample surface. A high bias voltage, between 150-180 V, was applied to the sample, which caused a field emission currents of 5-10 $\mu$A. In this STM operation mode,  the spatial (lateral) resolution is roughly similar to the tip-to-sample distance.\cite{Saenz1994} Therefore, the obtained CL spectra refer to a region of about one hundred nanometers wide of the monolayer h-BN.

\begin{acknowledgement}

This work was supported by the Fundação de Amparo à Pesquisa do Estado de São Paulo (FAPESP) Projects 14/23399-9 and 18/08543-7. This work at Nottingham was supported by the Engineering and Physical Sciences Research Council UK (Grant Numbers EP/K040243/1 and EP/P019080/1). We also thank the University of Nottingham Propulsion Futures Beacon for funding towards this research. PHB thanks the Leverhulme Trust for the award of a Research Fellowship (RF-2019-460). This work was financially supported by the network GaNeX (ANR-11-LABX-0014), the ZEOLIGHT project (ANR-19-CE08-0016), and the BONASPES project (ANR-19-CE30-0007).

\end{acknowledgement}

\section{Supplementary materials available }

AFM measurements and sample surface morphology. Tunneling parameters and STS. Electronic band gap determination and statistical analysis. Optical absorption. STM/STS of points defects. Correction in the luminescence spectra.

\bibliography{biblio}

\end{document}